\documentclass{jltp}

\usepackage{graphicx} 

\title{Leggett-Rice Systems in Harmonic Traps}
\author{R. Ragan$^a$ and J. Baggett$^b$}
\address{University of Wisconsin at La Crosse, La Crosse WI 54601 USA\\
$^a$Department of Physics\\
$^b$Department of Mathematics}
\runninghead{R. Ragan and J. Baggett}{Leggett-Rice Systems in Harmonic Traps}
\begin{document}
\maketitle
\begin{abstract}
We present two results concerning the spin (or pseudo spin) dynamics of trapped quantum gases in the hydrodynamic regime described by the Leggett equations. 
First, we apply perturbation theory to extend the ``bounded diffusion'' description to trapped systems for small field inhomogeneities. Second, we study the formation of long-lived domains with a numerical stability analysis of the lowest-lying longitudinal diffusive modes. We also use computer simulations of $\pi/2$-pulse experiments to determine the range of experimental parameters where the formation of domains can be observed.
\end{abstract}
\section{INTRODUCTION}      
Recent experiments at JILA\cite{1} and subsequent analyses\cite{2} have demonstrated the identical spin rotation effect (ISRE) in the pseudo spin dynamics of trapped ultra-cold atoms with two internal states. The ISRE in spin-polarized quantum gases\cite{3} and analogous mean-field effects in liquid helium\cite{4} give rise to interesting behavior such as spin waves, instability against transverse fluctuations, and spontaneous formation of magnetic domains.
Fuchs {\it et al.}\cite{5} have shown that simulations of the appropriate kinetic equations with JILA parameters reproduce the observed pseudo-spin wave oscillations.
%
%
Kuklov and Meyerovich\cite{6} have noted that Castaing instabilites\cite{7} and domain wall solutions should be present in trapped gases as well.
Recently, Fuchs {\it et al.}\cite{8} have investigated these instabilities by performing simulations of the kinetic equations with large initial longitudinal polarization gradients. Although the Castaing instability is washed out for low collision frequencies, they found that the stability should be observeable in trapped gases in, or near, the hydrodynamic regime. 

In this paper, we restrict our attention to the hydrodynamic regime (collision time $\tau \rightarrow 0$) where the equations of motion originally derived by Leggett and Rice\cite{9} (LR) for spin-polarized systems apply, and where the nonlinear dynamics are most evident.
In the first part, we analyze the behavior of trapped systems with small linear and quadratic field inhomogenities, although the results are readily generalized for arbitrary field inhomogeneities (and for finite $\tau$, for that matter). In the second part, we carry out a numerical stability analysis of the longitudinal diffusion modes, and use computer simulations to study the formation of domain walls in JILA-type $\pi/2$-pulse experiments.

For cigar-shaped traps situated along the z-axis the LR equations of motion reduce, in the Larmor frame of the center of the trap, to
\begin{equation}
\frac{\partial \vec{M}}{\partial t} =  G_{k}z^{k} \hat{z} \times \vec{M} 
- \frac{\partial \vec{J_{z}}}{\partial z} 
\label{dMdt}
\end{equation}
\begin{equation}
\vec{J}= -\left(D\frac{\partial M}{\partial z} + \Omega z M \right) \hat{e}- D \frac{M}{1+\mu^2 M^2}
\left( \frac{\partial \hat{e}}{\partial z}+ \mu M \hat{e} \times \frac{\partial \hat{e}}{\partial z}
\right)
\label{J}
\end{equation}
where $\vec{M}=M\hat{e}$, $\vec{J}$ is the current density, $G_{k}$ is the linear ($k=1$) or quadratic ($k=2$) strength of the field inhomogeneity, $D$ is the spin diffusion coefficient, $\mu M$ is the spin-rotation parameter, and $\Omega=\omega_{z}^{2}\tau$, where $\omega_{z}$ is the axial trap frequency. We do not take any polarization or density dependence of the transport coefficients into account.

These equations can be put into dimensionless form by writing $\Omega t \rightarrow t$, $z\sqrt{\Omega/D} \rightarrow z$, $\vec{m}\equiv \vec{M}/M_{0} $, $ \vec{j} \equiv \vec{J}/(M_{0}\sqrt{D\Omega})$, $\alpha \equiv \mu M_{0}$, $ g_{1}\equiv  G_{1}\sqrt{D/\Omega^{3}}$, and $ g_{2}\equiv D G_{2}/\Omega^{2}$, where $M_{0}$ is the equilibrium magnetization at the center of the trap. In spherical coordinates, we have
\begin{equation}
\dot{\vec{m}} =\dot{m} \hat{e}+m\dot{\theta}\hat{\theta}+m\sin{\theta}\dot{\phi}\hat{\phi}= g_{k}z^{k}  m\sin{\theta}\hat{\phi}  
- \vec{j}'
\label{dMdtsp}
\end{equation}
\begin{equation}
\vec{j}= -\left( m' + z m \right) \hat{e}-  \frac{m}{1+ \alpha^{2}m^2}
\left[ (\theta'-\alpha m\sin{\theta}\phi') \hat{\theta}+(\sin{\theta}\phi'+ \alpha m\theta')\hat{\phi} \right]
\label{Jsp}
\end{equation}
where dots (primes) denote temporal (spatial) partial derivatives. After a $\pi/2$ pulse the initial conditions are   
 $m = \exp(-z^2/2) $, $\theta = \pi/2$, and $\phi = 0$.

\section{SMALL FIELDS $(\alpha g_{k} \ll 1)$}
For small values of the parameter $\alpha g_{k}$ the system is in the bounded diffusion regime\cite{10} and the external field can be treated as a perturbation of the tipped equilibrium magnetization. 
In this case the field inhomogeneity gives rise to (1)  transients after the $\pi/2$-pulse that oscillate at the zero-field spin-wave frequencies and (2) a slow decay of the magnetization. The linearized equation of motion is obtained by substituting  $\theta=\pi/2+\delta \theta (t)$, $\phi= \delta \phi (t)$ into Eqs.(\ref{dMdtsp}-\ref{Jsp}), giving
\begin{equation}
m\dot{\psi}=\lambda m\psi =\left(\frac{m \psi'}{1-i\alpha m}  \right)'+ig_{k}z^{k}m \rightarrow \frac{i \psi''}{\alpha} +ig_{k}z^{k}m,\,\,{\rm as}\,\, \alpha \rightarrow\infty
\label{dpsi}
\end{equation}
where we have introduced the complex phase $\psi = \delta \theta(t) + i \delta \phi(t)$.
\begin{figure} 
 \centerline{\includegraphics[height=3.5in]{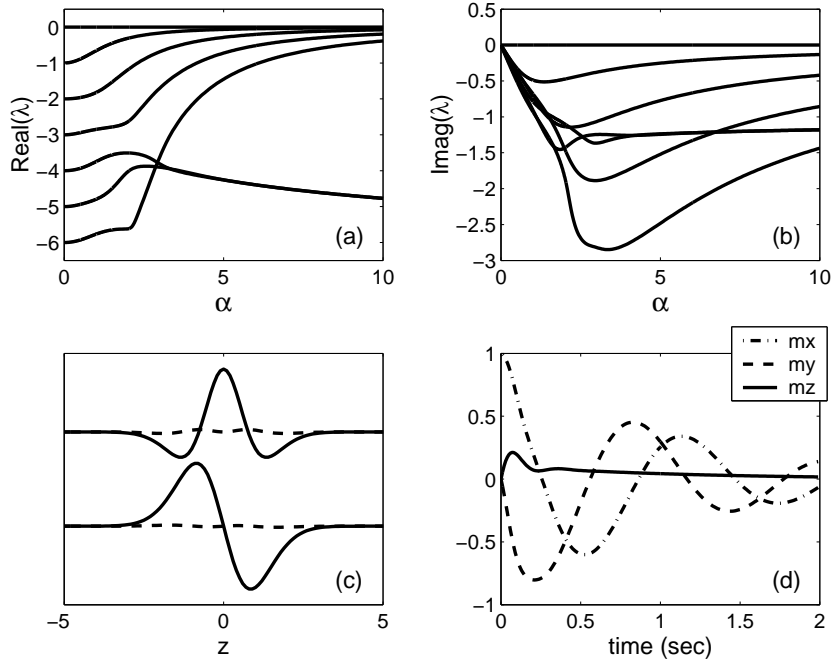}} 
 \caption{(a-b) The real and imaginary parts of the eigenvalue spectrum of Eq.\ref{dpsi}. (c) $n=1$ and $n=2$ eigenfunctions $m\psi_{n}$ for $\alpha\rightarrow \infty$, (d) Evolution of   $m_{z}$ at the center of the trap, using only the n=2 mode, for JILA parameters $n=1.8\times 10^{18} {\rm cm}^{-3}$, $\nu_{diff}=0.07 {\rm Hz}$ ($\alpha=4$, $g_{2}=0.25$). A bias Larmor frequency of $-2g_{2}$ ($-1.5 {\rm Hz}$) has been added to simulate a Gaussian field inhomogeneity.} 
 \label{fig:spect} 
 \end{figure}
The spectrum of the zero-field modes, shown in Figs. 1a-b, is obtained by setting $g_{k}=0$ and solving Eq.(\ref{dpsi}) with the boundary conditions $m \psi'\rightarrow 0$ as $z \rightarrow \infty$, where $m=\exp(-z^2/2)$.  For $\alpha=0$ the solutions are diffusive modes of alternating parity with $\lambda_{n}=-n$. For larger values of $\alpha$ some modes of opposite parity coalesce into modes localized on the left or right of the trap. These modes decay rapidly, however, and are not excited by small fields. Only the first two modes with eigenvalues $\lambda_{1}\approx -1.71/\alpha^{2}-1.32i/\alpha $ and  $\lambda_{2}\approx -7.65/\alpha^{2}-4.20i/\alpha $ are appreciably excited. 
The analysis is greatly simplified for large $\alpha$ in which case the RHS of Eq.(\ref{dpsi}) is Sturm-Liouville operator with kernel $m$.  To analyze the transients after a $\pi/2$ pulse, we write $\psi=\sum c_{j}(t)\psi_{j}$ and dot Eq.(\ref{dpsi}) with $\psi_{n}$, giving
\begin{equation}
\dot{c_{n}}=i\omega_{n}c_{n}+ ig_{k}b_{kn}
\end{equation}
where $b_{kn}=\int z^{k} m \psi_{n} dz$. This has the solution 
\begin{equation}
c_{n}(t)=\alpha g_{k}d_{kn}(\exp i\omega_{n}t -1)
\end{equation}
where $d_{kn}= b_{kn}/(\alpha\omega_{n})$. This solution is still valid for small damping with the replacement $i\omega_{n} \rightarrow \lambda_{n}$. 
After the transients the long-lived mode can be found from $\psi=\alpha g_{k}\sum_{n} d_{kn}\psi_{n}$. Alternatively, we set $j_{e}=0$ and $j_{\theta}=0$ in Eq.(\ref{Jsp}). This gives 
\begin{equation}
\theta'=\alpha m \sin{\theta} \phi' = \alpha m \phi'
\label{dtheta}
\end{equation}
Substituting this into Eqs.(\ref{dMdtsp}-\ref{Jsp}) gives equations of motion for the long-lived mode that are identical to the $\alpha=0$ case.  These can be solved exactly,\cite{11} and for small $g_{k}$ give $\delta\phi=\omega_{k0}t+g_{k}z^{k}$, where $\omega_{10}=0$ and $\omega_{20}=g_{2}$, with $m$ decaying at a rate $\lambda_{0}=g_{k}^{2}$. 

The $m_{z}$ profile is found by integrating Eq.(\ref{dtheta}) and using $\int m_{z}\, dz=0$. 
For linear and quadratic fields we get, respectively,
\begin{equation}
m_{z}=\alpha g_{1}\sqrt{\frac{\pi}{2}}{\rm erf}\left(\frac{z}{\sqrt{2}}\right)\exp{\left(-\frac{1}{2}z^{2}\right)}
\label{mzg1}
\end{equation} 
\begin{equation}
 m_{z}=\alpha g_{2}\left[\exp{\left(-\frac{1}{2}z^{2}\right)}-\frac{1}{\sqrt{2}}\right]\exp{\left(-\frac{1}{2}z^{2}\right)}
\label{mzg2}
\end{equation} 
which are essentially identical to the $n=1$ and $n=2$ zero-field modes (see Fig. 1c.). We note that for small $g_{k}$ and large $\alpha$ the largest corrections to the hydrodynamic results are $O(\omega_{n} \tau \propto \Omega\tau\/\alpha$). Fig. 1d shows the evolution  $m_{z}$ at the center of the trap, which is qualitatively similiar to the evolution for larger fields (compare to Fig. 1 in Ref. \onlinecite{5} where $\alpha g_{2}=4.8$) .
\section{LARGE FIELDS ($\alpha g_{k} \gg 1$)}
For larger values of $\alpha g_{k}$ the the modes are strongly coupled and higher-orders are evident in the transients. Even so, a linear analysis (which we do not provide here) involving the energy of the undamped modes correctly predicts $n_{max} \approx \sqrt{\alpha g_{k}}$ for the number of orders observed in simulations (see Fig. 2a). Also, for large $\alpha g_{k}$ we can treat the trap as a perturbation for short times, giving $t\sim (\alpha/g_{k})^{1/2}$ as the time for the
boundary conditions to propagate to the center of the trap (i.e. $|z|\ll1$). Before this time the magnetization at the center of the trap is effectively unbounded, and for large $\alpha$ evolves according to $\phi=g_{1}zt$, $\theta = \pi/2$  ($k=1$); or  $\phi=g_{2}z^{2}t$, $\theta = \pi/2+4g_{2}^{2}t^{2}/\alpha$ ($k=2$).

%
For large values of $\alpha g_{k}$ a $\pi/2$ pulse can result in the formation of a long-lived domains. These structures, which have been extensively studied in liquid helium,$^{12,13}$ are formed as the z-component of the spin current $j_{z}$ drives the dipole (k=1) or quadrapole (k=2) longitudinal modes.
\begin{figure} 
 \centerline{\includegraphics[width=5in]{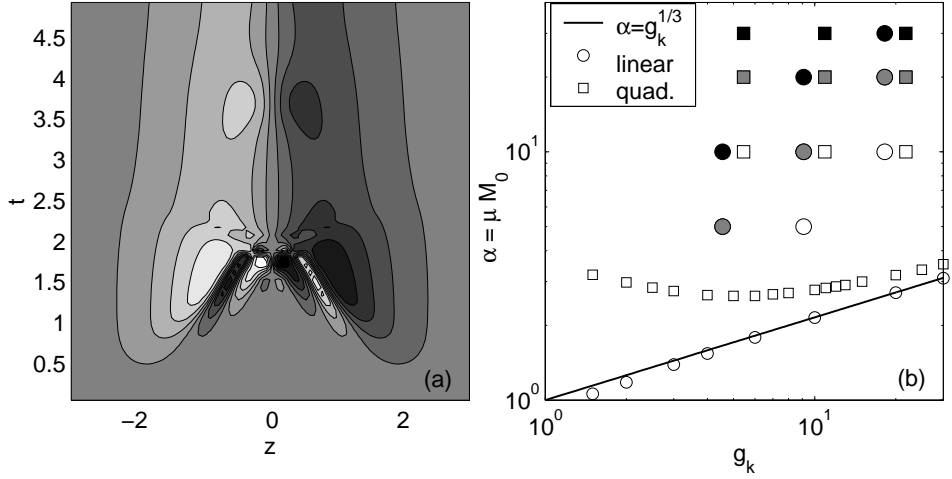}} 
 \caption{(a) Evolution of $m_{z}(z,t)$ for $\alpha=15$ and $g_{1}=5$ showing the higher-order transients and the formation of a domain wall. (b) $g_{k}-\alpha$ stability diagram. The small symbols show the results of the numerical stability analysis of the $n=1$ and $n=2$ longitudinal modes. The larger symbols indicate whether a domain wall forms (black), or not(white), (gray=marginal), in simulations of $\pi/2$-pulse experiments.} 
 \label{fig:wall} 
 \end{figure}
As the transients decay, the transverse magnetization ($m^{+} \equiv m_{x}+im_{y}$) becomes confined near the zeroes of $m_{z}$. Normally, the field gradient dephases $m^{+}$ rapidly and $m_{z}$ decays via ordinary diffusion.  However, if $\alpha$ is large enough, the longitudinal spin current is unstable, and $m^{+}$ develops into coherent domain walls, dividing $m_{z}$ into two (k=1) or three (k=2) domains that decay via transverse diffusion across the domain walls. For large $\alpha g_{k}$
the domain wall solution is given by\cite{12}
$\theta''=\alpha(\omega_{k}-g_{k}z^{k})\sin{\theta}$,  
where $\omega_{1}=0$ and $\omega_{2}=g_{2}z_{0}^{2}$, and $z_{0}$ is
a zero of the z-magnetization given by ${\rm erf}(z_{0}/\sqrt{2})=1/2$.  This description is valid so long as the width of the domain wall $\approx (\alpha
g_{k})^{-1/3} \ll 1$.

For large gradients the walls are destroyed
for $g_{k}>\alpha^{3}$ as found in Ref. \onlinecite{14} for uniform $j_{z}$. To find the stability threshold for smaller fields, 
we performed a numerical stability analysis of the $n=1$ ($m_{z}=zm$) and $n=2$ ($m_{z}=(1-z^{2})m$)
longitudinal modes against transverse perturbations. There is a
complication that the longitudinal modes decay at rates $\lambda_{n}=-n$.
For a given $g_{k}$, we defined the threshold of stability of the $n$th mode as the value of $\alpha$ where
the perturbations grow at the rate $+n$. The results
are summarized in Fig. 2b. 
Determining the minimum $\alpha$ for the formation
of domain wall after a $\pi/2$ pulse is difficult because the
magnetization decays in a rather complicated manner before the longitudinal modes are established. Thus, we carried out simulations of $\pi/2$
pulse experiments for various values of $\alpha$ and $g_{k}$. The
presence of a domain wall is signaled by its slow decay rate\cite{12} $ \lambda \approx \alpha^{-2} \int m^{-1}(\theta')^{2}dz \approx 2.5 g^{1/3}/\alpha^{5/3}$ that is much slower than the longitudinal diffusion rate. The results, which are summarized in Fig. 2b, indicate that domains will not be observeable in quadratic (or Gaussian) fields for $\alpha< 20$. However, for linear fields (perhaps in optical traps) domains should be observable for $\alpha \sim 10$. 

Recently, domains have been observed \cite{15} in normal trapped $^{87}$Rb, but since the lifetime of the domains was comparable to their formation time, the experiment corresponds to the ($n=2$) marginal category in Fig. 2b. 
\section*{ACKNOWLEDGEMENTS}
This research was supported by NSF Grant DMR-0209606.
%

\end{document}